\documentclass[smallextended]{svjour3}       
\smartqed  
\usepackage{graphicx}
\usepackage{hyperref}
\usepackage{gensymb,siunitx} 
\usepackage{booktabs}
\usepackage{cite}
\usepackage{lipsum}
\usepackage{mathtools}
\usepackage{geometry}
\usepackage{ braket,  array}

\begin{document}

\title{Quantum Communication using Code Division Multiple Access Network
}


\author{Vishal Sharma         \and
        Subhashish Banerjee 
}


\institute{Vishal Sharma \at
                  IIT Jodhpur, Rajasthan, India.\\ 
                  \email{pg201383506@iitj.ac.in}.          
           \and
          Subhashish Banerjee \at
          IIT Jodhpur, Rajasthan, India.\\
            \email{subhashish@iitj.ac.in}.  
}

\date{Received: date / Accepted: date}

\maketitle

\begin{abstract}
	For combining different single photon channels into single path, we use an effective and reliable technique which is known as quantum multiple access. We take advantage of an add-drop multiplexer capable of pushing and withdrawing a single photon into an optical fiber cable which carries  quantum bits from multiusers.  In addition to this, spreading spreads the channel noise at receiver side and use of filters stop the overlapping of adjacent channels, which helps in reducing the noise level and  improved signal-to-noise ratio. In this way, we obtain enhanced performance of code division multiple access-based QKD links with a single photon without necessity of amplifiers and modulators. 

\end{abstract}

\section{COMMUNICATION CHANNEL WITH MULTIPLE USERS}
\label{intro}

The successful transmission of classical data is achieved by various quantum key distribution \cite{bennett1984quantum, boaron2018secure,  ekert1991quantum, sharma2019analysis, sharma2016comparative, sharma2016effect, sharma2018analysis, thapliyal2017quantum}, and other quantum communication protocols \cite{bennett1992communication, bennett1993teleporting, sharma2015controlled}. In most of these protocols, we prefer photons as information carrier through various optical fiber links \cite{gisin2002quantum, hiskett2006long,olmschenk2009quantum, townsend1994quantum, hughes2000quantum, takesue2015quantum}.\newline

In case of high traffic density, where more than one user want to access the channel at the same time, multiple access schemes \cite{sklar1983structured, sklar2001digital, benslama2016transitions, belavkin2013quantum} comes into picture to fulfill the demand of channel access at the same time. As far as quantum networks are concerned, wavelength and frequency division multiple access schemes play an effective role in which each user transmits information at different frequencies  \cite{brassard2003multiuser,yoshino2012high, chapuran2009optical, qi2010feasibility,  tanaka2008ultra, patel2014quantum, ortigosa2006subcarrier, yoshino2012high, mora2012experimental, ciurana2014quantum, eriksson2019wavelength}, and each user wait for their chance, in case of time division multiple access technique \cite{choi2011quantum}.\newline

Here, we are using concepts of multiple access which is based on spread-spectrum methods \cite{sharma2014analysis}. Most of our discussions are based on optical fiber based data transmission, but it could be useful for free-space communication. In our scheme we are deploying optical devices, which is very useful for transmitting photons of multiple users via the same optical fiber. In this way, the multiple users share their time window, path, and frequency band. Hence, this quantum optical communication approach is based on spread-spectrum multiple access methods, and eliminates heavy losses, which were present in previous schemes \cite{razavi2012multiple, zhang2013quantum}. In our scheme, along with add-drop architecture, we are adopting classical spread-spectrum techniques \cite{sharma2014analysis}.

	  	    	   

\section{SPREAD SPECTRUM TECHNOLOGY}

In spread spectrum techniques \cite{sharma2014analysis}, bandwidth $B$ of a modulated signal $D(t)$ is spread by $RB$, where $R$ is the spreading factor \cite{pickholtz1982theory}. The received signal $S(t)$, after spreading operation, is a larger bandwidth signal, as compared to the original modulated data signal $D(t)$.  \newline

In our current work, we are using DSSS (Direct sequence spread-spectrum) method, in the CDMA (Code division multiple access) technology. In CDMA, a code $c_{p}$ ($1$ or $-1$ as a vector) is assigned to each user $U_{p}$. The basic and essential condition while assigning $c_{p}$ to each user is that, $c_{p}.c_{q}^{T} = \delta_{pq}$, or $c_{p}.c_{q}^{T}\leq r$, for $p \neq q$, where value of integer $r$ must be as small as possible. \newline


During spreading process, signal is multiplied by $c_{p}$. Despreading can be achieved at the receiver side by again multiplying the signal by the same value of $c_{p}$.\newline

To separate the signals from all the users, who share the bandwidth at the same time, it is highly essential that the codes  be chosen appropriately. The operations of spreading diminishes the effect of noise and permits to enhance the multiple users with improved separation. In this task, an appropriate length of orthogonal codes are chosen. \newline
 
DSSS (Direct-sequence spread spectrum) techniques are employed to single photons \cite{belthangady2010hiding}. First we spread the photon's wave function and then despread at the receiver side. In despreading operation, we get back the photon's wavefunction in original form. If some noise exists, filters are used to remove the noise part from photon's wavefunction.\newline

In this work, we present advantages of using single photon spreading for photon's channels where improved performance is obtained alongwith multiple access systems. 

\section{BASIC BUILDING BLOCKS}
 
 We are incorporating three important optical fiber elements as the basic building blocks of the working system namely: fiber Bragg gratings (FBG), circulators, and modulators. A signal's wavefunction is altered by a control signal \cite{saleh1991fundamentals},  present in the electro-optic modulators. In the same context, we can follow the same approach in the quantum domain \cite{capmany2010quantum}. The photon's phase are modulated using optical modulators. For this, we can apply \cite{belthangady2010hiding},  for a phase shift of $\pm \frac{\pi}{2}$ to various time bins of the photon's wave function. In this procedure, we use a wavefunction of time length $T$ to segment it into $S$ different segments. It is the corresponding code element $C_{p}$ which approves phase shift $\frac{\pi}{2}$ (if the element is 1), or -$\frac{\pi}{2}$ (if the element is -1). In addition to this, to make phase 0 (for element 1) or $\pi$ (for element -1), we use an additional optical modulator. At the end, the total phase change by both the modulations are $\pm \frac{\pi}{2}$ or 0/$\pi$.\newline
 
\begin{figure}[h]
\centering
\includegraphics[width=0.70\textwidth]{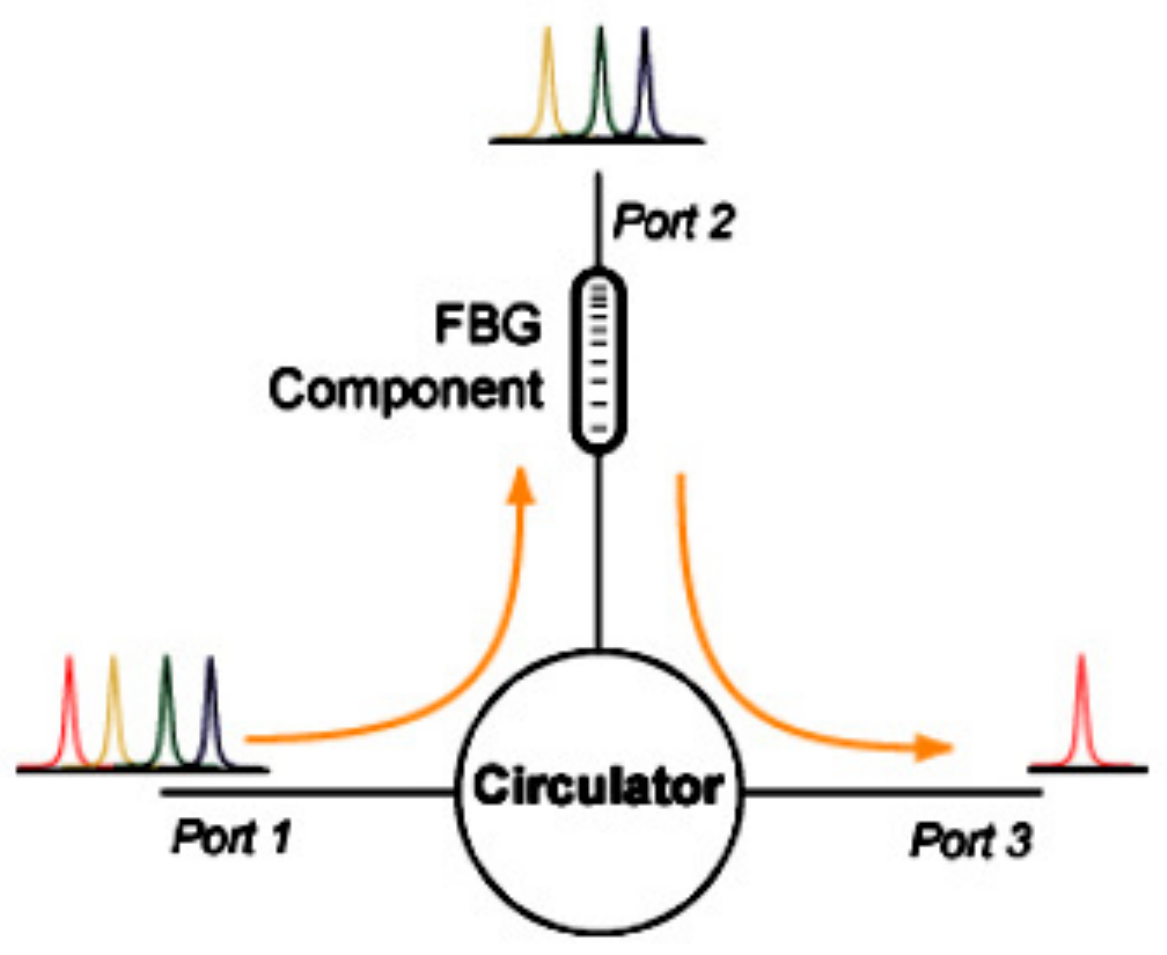}
\caption[1]{ Circulator operation with Fiber Bragg Grating (FBG)\cite{thorlabs1}.}  \label{circulator1} 
\end{figure}

\begin{figure}[h]
\centering
\includegraphics[width=0.800\textwidth]{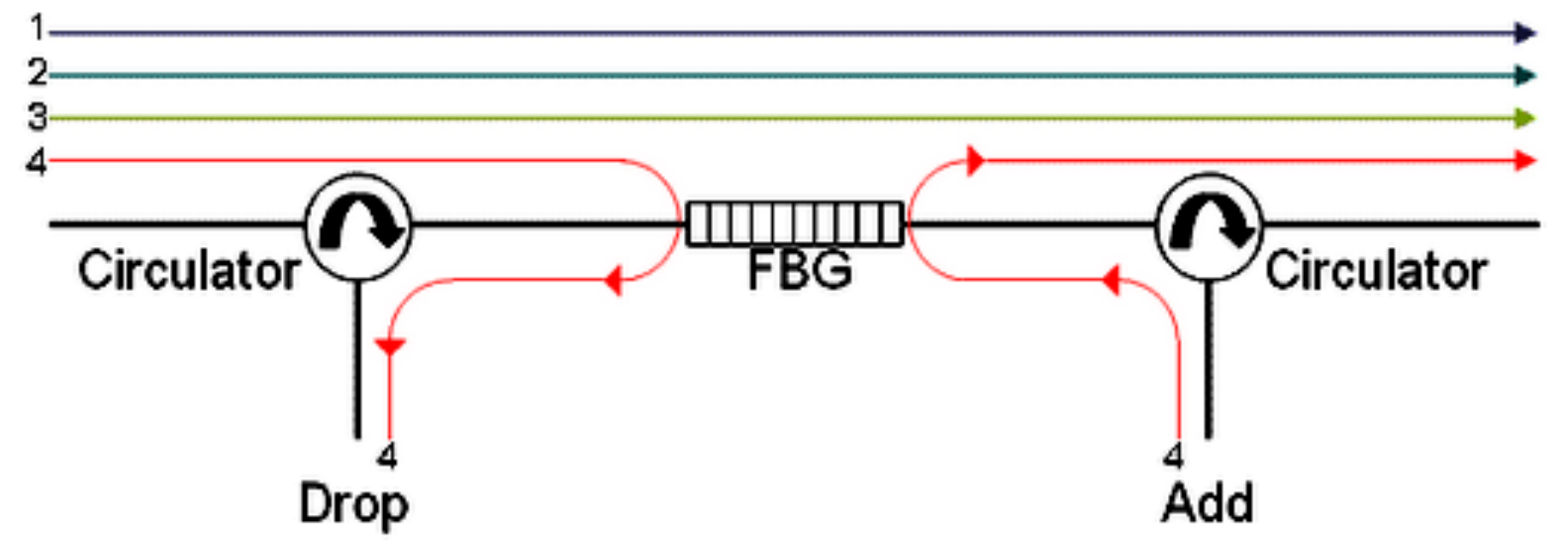}
\caption[1]{Optical add-drop multiplexer (OADM) with a fiber Bragg grating and two circulators \cite{thorlabs2}.} \label{circulator2} 
\end{figure} 

We now require a technique to add the spread photons of  multiusers in the same fiber. We are using two important optical elements: fiber Bragg gratings and circulators. Circulators are non-reciprocal optical devices that redirect incoming optical signals to the successive output ports (as shown in Figs. \ref{circulator1} and \ref{circulator2}). \newline

To increase the bandwidth over optical fiber transmission, a multiplexing technology known as Dense Wavelength Division Multiplexing (DWDM) is used. This is an optical multiplexing approach, which transmits various signals at different wavelengths on the same fiber.\newline

Optical circulators are three-port devices which are used in a wide range of optical setups. These are non-reciprocating and polarization-maintaining (PM) optical circulators  with a center wavelength of 1310 nm, 1550 nm, and 1064 nm. The optical circulators have many significant properties which are required in communication systems such as very low insertion loss and high isolation. Because of these unique properties, these are used as chromatic dispersion compensation devices, add-drop multiplexers, and  bi-directional pumps.\newline


 Fig. \ref{circulator1} describes the operation of circulators with Fiber Bragg Grating (FBG). Circulators drop an optical signal from a dense wavelength division multiplexing (DWDM) technique with the use of FBG. The Port 1 is coupled to the input of DWDM with Port 2 connected to FBG. The FBG reflects the single wavelength, which then reenters the circulator in Port 2  and finally reaches at Port 3. The rest of the signals travels through FBG and reaches at the top fiber. \newline

\begin{figure}[h]
\centering
\includegraphics[width=0.800\textwidth]{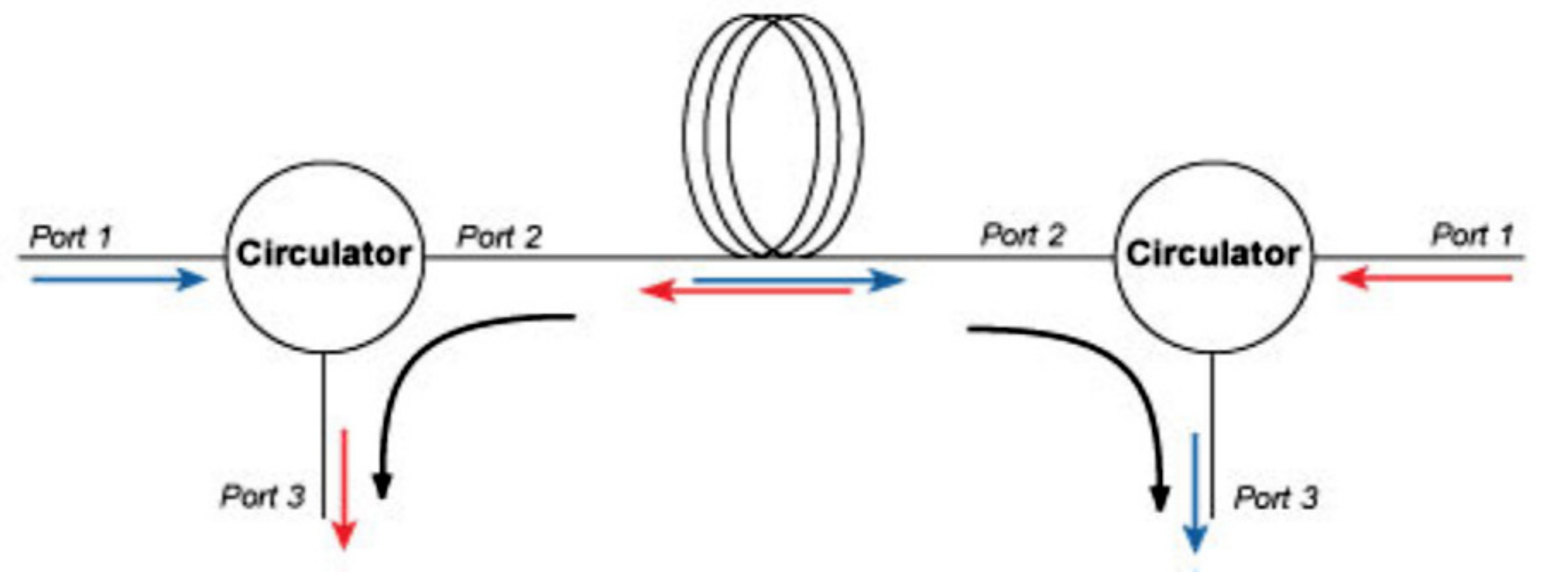}
\caption[1]{Circulator operation at both the ends of the fiber \cite{thorlabs3}.} \label{circulator3} 
\end{figure}

 Fig. \ref{circulator3} explains the operation of the two circulators deployed at the end of the fiber, which add the signals in one direction and remove the signals in the other direction. Hence, this kind of arrangement is used to transmit optical signals in two different directions down a single fiber.\newline

\begin{figure}[h]
	  	       		    \centering
	  	       		     \includegraphics[width=0.800\textwidth]{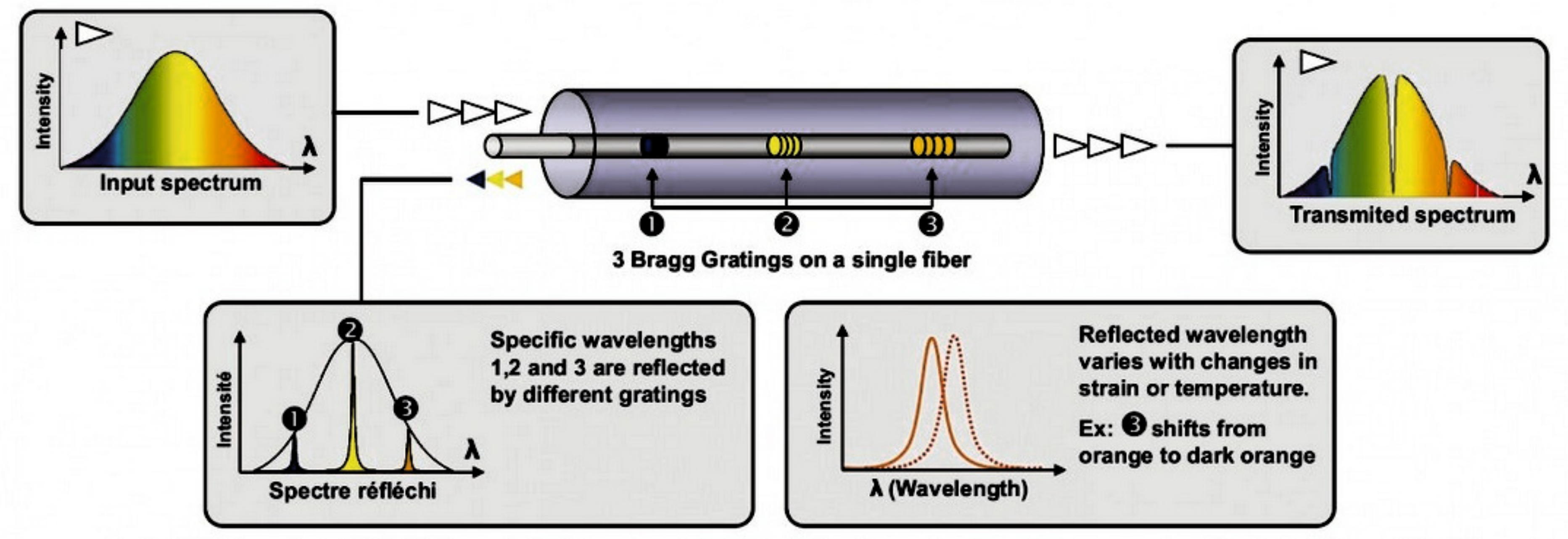}

	  	       				  	       		  	    	    	  	  	    \caption[1]{Operation of Fibre-bragg-grating \cite{scaime}.}  \label{fibergrating} 
 		  	    	    	  	  	    	    	  	 	    	    	  	  	    	    	    \end{figure}

Fiber Bragg gratings allow to pass many of the incoming signals without altering their properties, in case when a special frequency band is reflected. Hence, FBGs are called reflectors for some specific frequencies (as shown in Fig. \ref{fibergrating}).

\section{SIGNALS FOR QUANTUM COMMUNICATION}

We make use of qubits as information carrier with different encodings. In classical communication, out of different modulation schemes, frequency modulation is one of the most common approach to deal with such kind of transmissions. Following the same approach in quantum regime, $|0\rangle$  and $|1\rangle$ can be considered as wavefunctions at different frequencies \cite{capmany2012realization}. In addition to this, for getting phase information, at sidebands of carrier, we can use phase modulation \cite{guerreau2003long}.\newline

Time-bin encoding is also an alternative approach where wavefunction is bounded within (0, $T_{0}$) for encoding the quantum state $|0\rangle$, and $|1\rangle$ is encoded by introducing delay in wavefunction to start from ($T_{0}$, $2T_{0}$) \cite{brendel1999pulsed}. Time-bin quantum bits transmission scheme is mostly used in quantum key distribution and optical fiber communication. In time-bin qubit encoding scheme, codes are designed for a fixed time interval say $T_{0}$. After code design, these are applied twice within that time interval $T_{0}$, so that interference between the generated spread signals by different users can be  avoided. In this way, two orthogonal codes which produce similar spread signals will never overlap with each other and effect of the noise is diminished.

\section{MULTIPLEXING OPERATION}
\begin{figure}[h]
	  	       		    \centering
	  	       		     \includegraphics[width=0.800\textwidth]{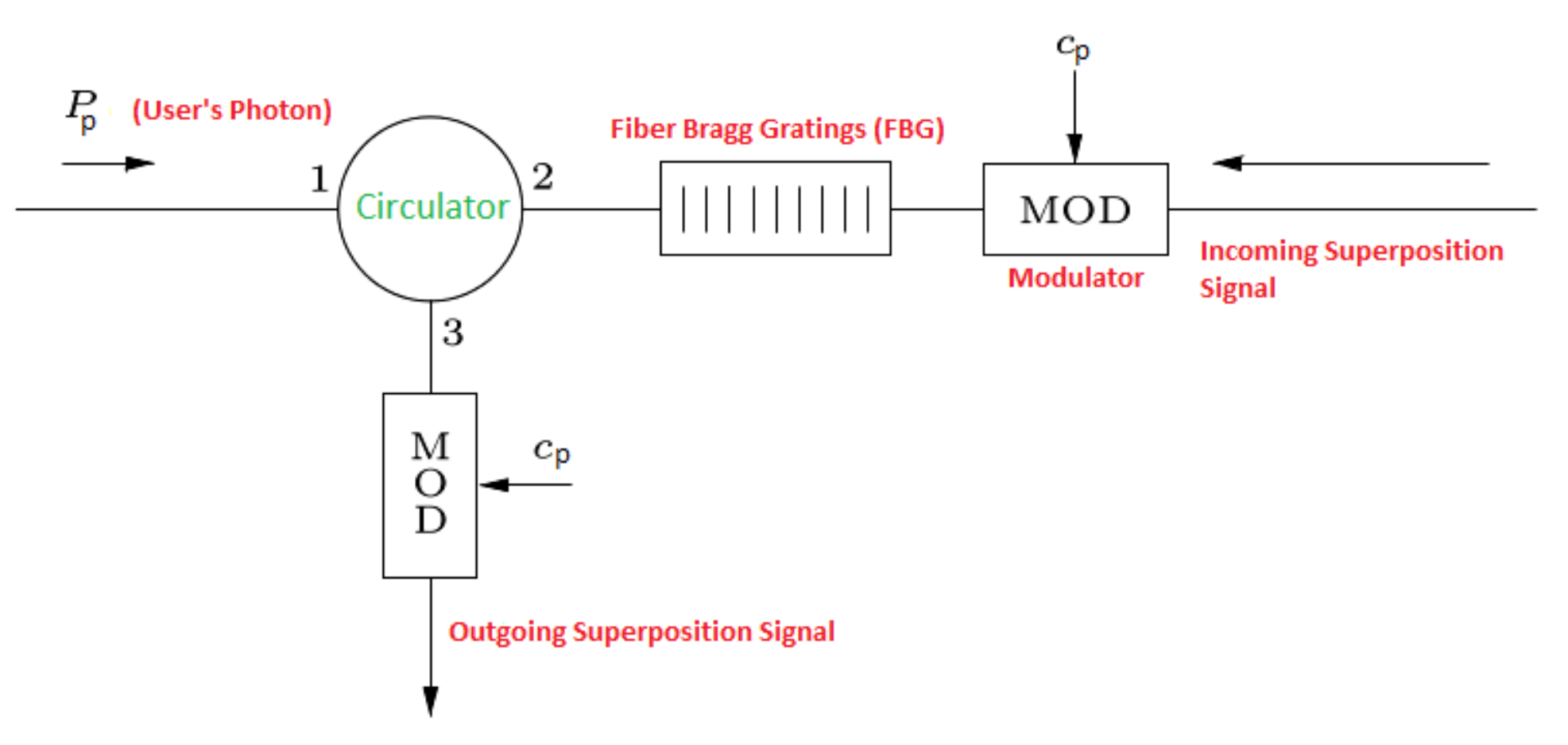}

	  	       				  	       		  	    	    	  	  	    \caption[1]{Multiplexer: The user's Photon ($P_{p}$) and incoming superposition signal meet at FBG. Here,  $c_{p}$ is the chipping code, which is used to multiply with incoming superposition signal. } \label{mux} 
 		  	    	    	  	  	    	    	  	 	    	    	  	  	    	    	    \end{figure}

In this section, we will discuss transmitter unit of each user which combine incoming signals with the $|\phi_{s}\rangle$. This $|\phi_{s}\rangle$ is a superposition state, and it has all the photons of the previous users. The multiplexing operation is shown in Fig. \ref{mux}. The add-drop multiplexing is used in optical fiber which helps in combining different frequencies from different channels. In our current work, we are using add-drop multiplexers in both the operations of multiplexing and demultiplexing.\newline

An optical add-drop multiplexer (OADM) is used for multiplexing light in single mode fiber (SMF). This is useful in wavelength-division multiplexing and in routing various channels of light into or out of a SMF. For designing various   
optical telecommunication networks, OADM acts as an optical node. Add and Drop terms are used here to add and remove the wavelengths according to the desired operations to take place during the transmission in a particular network. Hence, OADM is also known as an optical cross-connect.\newline

For adding individual photons of different quantum states, classical methods fail. The main reason is that classical methods doesn't follow the reversible operation of quantum states. Y-junction, star couplers are some well-known classical methods but they introduces high losses and fail to maintain quantum coherence \cite{razavi2012multiple, zhang2013quantum}, \cite{salehi1989code, fouli2007ocdma}. In our current add-drop multiplexer technique, photon loss is minimized. \newline

In Fig. \ref{mux}, the multiplexing operation is shown. The code $c_{p}$ is multiplied to the input superposition $U_{p}$. This multiplied signal is pass through fiber Bragg grating (FBG). The main function of FBG is to reflect those frequency bands which have the photon signal. This operation of frequency reflection is achieved before the spreading operation. \newline

Along with the above operations, we transmit the modulated signals in opposite direction without spreading, $d_{p}(t)$, as a data signal of next user via a circulator which reaches FBG simultaneously as a superposition signal. The new photon and other spreaded superposition signals are present in the circulator, where some of the signals come from FBG. The new photons are reflected by grating into port 2 and some part of the spread superposition return to the optical fiber it arrives from. \newline

The signal with previous photons and incoming new photon arrives the circulator, and is pointed to the next modulator in port 3.\newline

Here, at port 3, one more code $c_{p}$ is used to multiply. At this stage, photon from $U_{p}$ is spread with the input superposition to it's original form. \newline

The final result is the combination of the previous wavefunction, and the incoming new photon. In principle, the photons are less interactive with the external environment, as well as they don't interact with each other. Hence, this final result is similar to the tensor product of orthogonal wavefunction which lies in the same frequency range. \newline

Fiber bragg grating reflects some part of the photon's wavefunction. In case of S (spreading factor) is larger, the probability, $\frac{1}{S}$, which is reflected back, and lost will be small. This small value indicates very low value of channel loss. Further, for multiusers, this loss limits the channel access. For security reasons, different codes are used to spread the wavefunction at each multiplexer unit. The section of the wavefunction, which is in the spectrum as an ending part, is different for every extra photon added. We can think this as the resulting effect as a uniform loss, which is the effect of the second modulation.

\section{DEMULTIPLEXING METHOD}

\begin{figure}[h]
\centering
\includegraphics[width=1.000\textwidth]{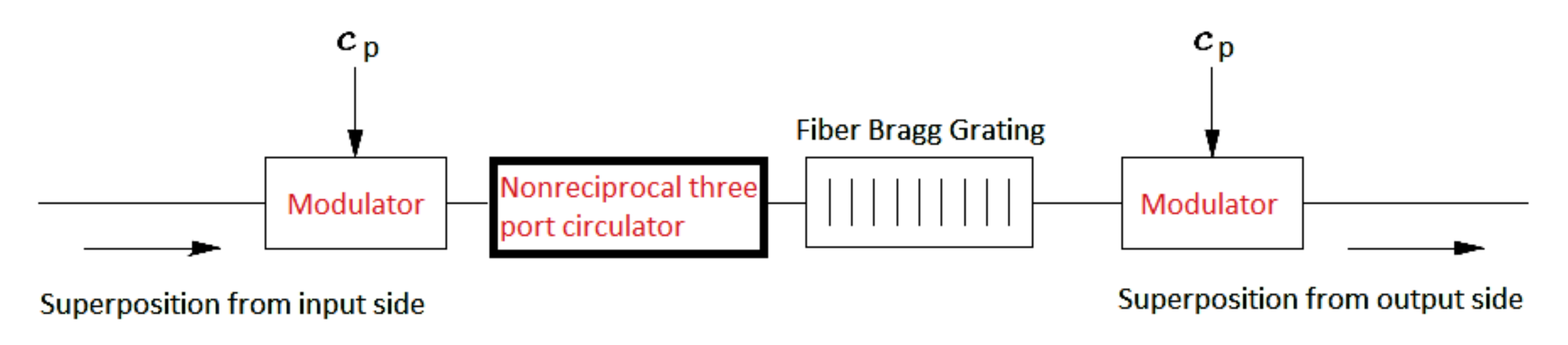}
\caption[1]{Demultiplexeing operation at the receiver end. } \label{demux} 
\end{figure}

If we perform the multiplexing operation for the $N$ users, we get qubits from all the users, where a fiber carries $N$ photons. At the receivers end, we separate out these photons using demultiplexing operation, as shown in Fig. \ref{demux}. As the circulator is a non-reciprocal device, this is the reason that we cannot use multiplexing circuit in reverse order in the demultiplexing operation. Hence, we have to perform slight modification in circuit sequence.\newline



As shown in Fig. \ref{demux}, the incoming superposition signal is modulated with the help of a code $c_{p}$, which results in spreading of the signal. This  spreaded signal is now despreaded by the modulator at the other end.  The final wavefunction is concentrated in the original spectrum $W$. \newline

The new signal is further sent  into an Fiber Bragg Grating. The FBG reflects the intended photon back to that circulator which forwards it to its intended receiver. In this process, some part of the noise is added in the form of a fraction $\frac{1}{S}$ of the wavefunction of the other photons. To recover all the incoming photons, we need to run the experiment N times. \newline

\section{SIMULATION FOR MULTI-USERS}

Photon losses become larger as compared to ideal case, where losses are $\frac{1}{S}$. This condition arises when we go through the practical implementation of the present study. Here, we show the extra benefits of the add-drop architecture in quantum spread spectrum. To nullify the effects of losses, and crosstalk between users, we use long length codes for obtaining better results.

\subsection{Photon Shape and their analysis}

Here we performed multiplexing operation on single photons of length $T$, which are in time domain. These single photons can carry an empty or a Gaussian wave packet. These time bin encoding schemes are similar to the COW (coherent one-way) protocol \cite{stucki2005fast, grosshans2003quantum, stucki2009continuous}, and Ekert protocol \cite{ekert1991quantum}, which are entangled in time-energy form \cite{tittel2000quantum}. The two same length time bins exist in a COW protocols, with a mean photon smaller than one. Here the Gaussian pulse is used to denote the time bin in the coherent one-way protocol. Out of the two time bins, the first and the second are corresponds to $|0\rangle$ and $|1\rangle$.\newline

We apply the multiplexing operation in a time-energy quantum key distribution protocol. This multiplexing technique helps in the  distribution of entangled pairs,  which were initially shared between each of the communicating users. Specially,  in a time-energy QKD method we allow the entangled photons to be shared between the authenticate users, these entangled photons were initially emitted from a significant process, known as spontaneous parametric down-conversion \cite{fedorov2009gaussian}. Continuous-variable (CV) quantum key distribution is not similar from the standard QKD  in the approach for detecting weak optical signals \cite{hirano2003quantum, grosshans2003quantum, jouguet2013experimental, wang201525, huang2016long, imre2012advanced, hanzo2012wireless}. \newline

The Gaussian wave packets are used to analyze the results. These Gaussian packets are set with $\sigma$ (standard deviation) = 0.1 T, with a centered time bin. We analyzed our results with in phase and random phase wave packets. Both these patterns of in phase and random phase are similar.\newline

Let $\phi(x, t)$ is the photon's wave function. This is a complex valued function of two real variables $x$ and $t$. If the wave function is considered as a probability amplitude, the square modulus of the wave function will be a positive real number: $$|\phi(x, t)|^2 = \phi^*(x, t)\phi(x, t) = \rho(x, t).$$ The asterik ($*$) denotes the complex conjugate. If the particle's position is measured, it's location cannot be determined from the wave function, but it is determined by a probability distribution. The probability that its position $x$ will be in the interval a $\leq$ x $\leq$ b is the integral of the density over this interval 
\begin{eqnarray}
P_{a \leq x \leq b}(t) = \int_{a}^{b} |\phi(x, t)|^2 dx,
\end{eqnarray} \label{pdf}
 where $t$ is the time at which the particle was measured. This leads to the normalization condition
\begin{eqnarray}
\int_{-\infty}^{\infty} |\phi(x, t)|^2 dx =1,
\end{eqnarray} \label{normalize}
because the particle is measured, there will be 100$\%$ probability that it will be somewhere.\newline
In our work, we have considered photon pulse as a Gaussian wave packet. The normalized  wave packet centered on $x = x_{0}$, and of characteristic width $\sigma$ is $$\phi(x) = \phi_{0}e^{-(x - x_{0})^{2}}/(4\sigma^{2}).$$ The normalization constant $\phi_{0}$ is  $$|\phi_{0}|^{2} = (1/2\pi\sigma^2)^{1/2}. $$ Hence, the  general form of normalized Gaussian wavefunction is 
\begin{eqnarray} 
\phi(x) = \frac{e^{i\varphi}}{(2\pi\sigma^{2})^{1/4}}e^{-(x-x_{0})^{2}}/(4\sigma^{2}),
\end{eqnarray} \label{Gaussian}
where $\varphi$ is an arbitrary real phase-angle. This is used for calculating the photon loss  and crosstalk probabilities in the coming sections.

\subsection{Codes for Multiplexing-Demultiplexing operations}

The linear feedback shift registers (LFSR) are used to generate a unique code for each user which is used in the spread spectrum techniques. A long sequence of pseudo-random binary sequences are desired for high speed data transfer as well as to scramble the message signals. The Pseudo noise (PN) sequences are generated by Exclusive-OR (X-OR) gates and shift registers, depends on the value of n (number of registers) and proper feedback, $2^n - 1$, as a periodic output is received, which is also known as maximum-length sequence ($m$-sequence). These m-sequences follow particular type of properties. These $m$-sequences generate random signal statistics with a low value of the autocorrelation function \cite{golomb2005signal}, a parameter to measure the similarity or dissimilarity between the input and output data. In a specific case, to analyze our results for performance evaluation, the shifted version of the various m-sequences with the right feedback can be chosen from the table in \cite{mutagi1996pseudo}. The code value $c_{p}$ denotes a cyclic shift by $i^{th}$ position. Following these codes from 1 to $2^n - 1$ with binary data as $\pm1$, the expression, $c_{p}.c_{q}^T  = -1$, is valid if and only if, p  $\neq$ q. In other case, the expression $c_{p}.c_{q}^T$ = $2^n - 1$, is valid if and only if, both the indices are equal, i.e., $p = q$. \newline

 We need proper synchronization between the modulators. In addition to this, deploying a classical side channel can coordinate the nodes, which help in maintaining the start and end timings of the bit stream, and also codes will be orthogonal. These issues are discussed in QKD networks \cite{stucki2005fast}, \cite{tanaka2008ultra}. Without proper synchronization between the multiplexer and demultiplexer units many undesirable effects such as interference and change in code families may occur, which further leads to different sensitivities to time shifts.

\subsection{Noise removing filters and signal modulators}

In our study, we are considering FBG filter of spectral width $\sigma_{filt}$. This is considered similar to the Gaussian filter. If we compare this spectral width  with the spectral width of the Gaussian wave packet of the photons, we find that it is $\frac{8\pi}{5}$ times the spectral width of the FBG filter. Specially for FBG, we have chosen Gaussian shapes, the reason being that, the properties of the Gaussian shapes are quite suitable and come close to the transfer function of apodized gratings. \newline

We are using an ideal modulator, which provides  a 0 or a $\pi$ phase shift. In the interval of length $\frac{T}{S}$, the modulator introduces variable phase shift, which are determined by the binary values of the code used. These code elements are generated at regular intervals of $T$ seconds. We receive $S$ elements after $T$ seconds. There will be no phase shift introduced by the modulator, in the case the code element is 1 and if the code is -1, $\pi$ shift occurs.

\subsection{Sources for insertion Losses and Noise }

In the simulation, we assume the circulators, modulators and all the connections are ideal and have no losses. The aim of this simulation is to model intrinsic losses due to the spreading. The effect of other sources of loss is commented in the discussion. Likewise, we have not included noise or other effects that can degrade the signal.

\subsection{Simulation results}

In this system, we receive two different sources of errors: one source of error due to crosstalk and other due to photon loss. These values vary with the value of spreading factor ($S$) and the number of users. Our results represent that overall performance can be enhanced with larger value of spreading factor, but as the number of users participate in the communication process, the channel performance degrades. Hence larger value of spreading factor helps in increased system performance, but each additional user degrades the system performance. It is a trade off between the two factors.\newline

\begin{figure}[h]
	  	       		    \centering
	  	       		     \includegraphics[width=1.10\textwidth]{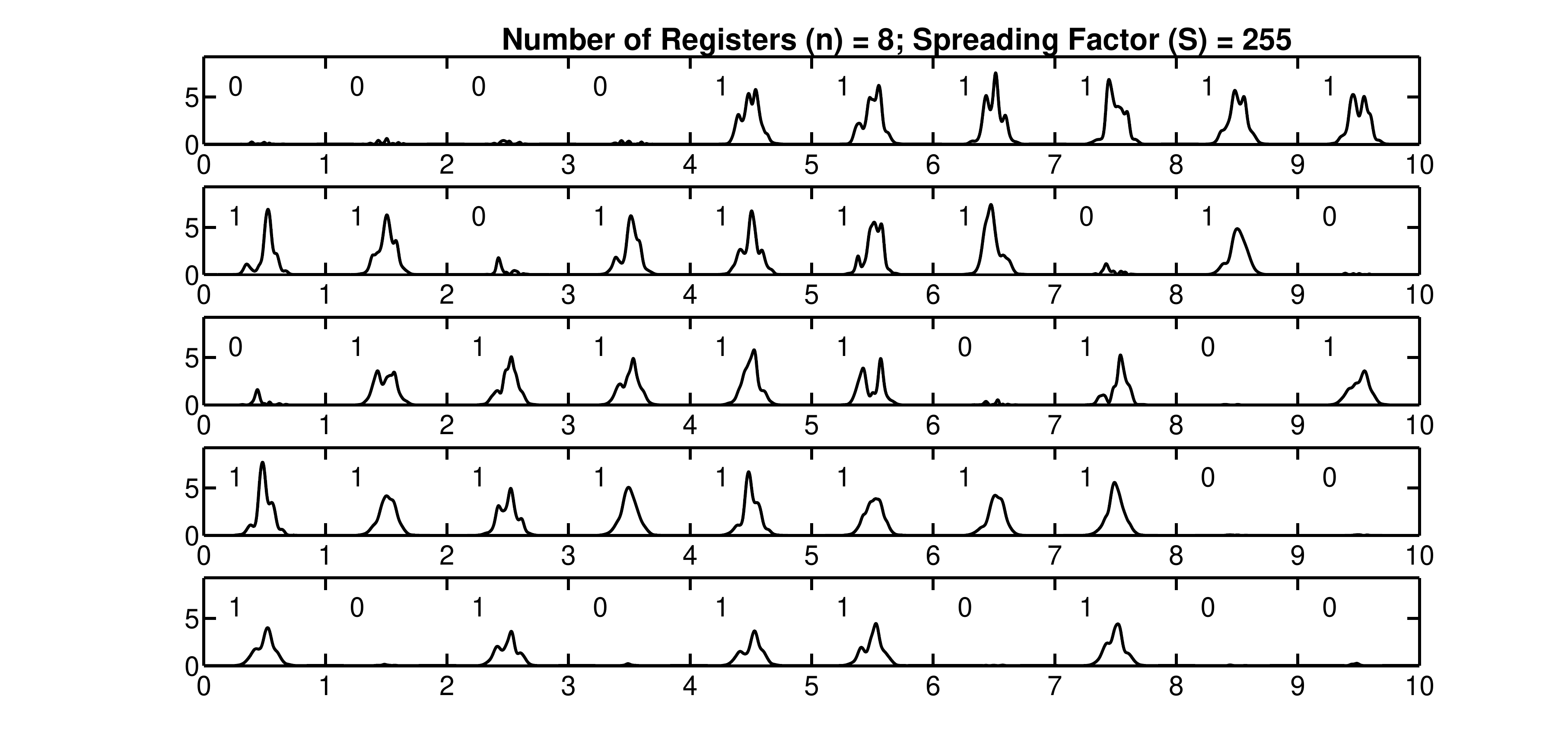}

	  	       				  	       		  	    	    	  	  	    \caption[1]{Spreading factor, $S$ = $2^{8}$ - 1 (with a code from an LFSR (Linear Feedback Shift Register) and n = 8 registers) for five users. The figure shows the output in terms of the density of the average photon number. Results show losses (low amplitude 1 pulses) and crosstalk (pulses in 0 bins and  high amplitude pulses in 1 bins). In an ideal system, the represented output corresponds to the probability density squared $|\phi(t)|^2$. } \label{n80} 
 		  	    	    	  	  	    	    	  	 	    	    	  	  	    	    	    \end{figure}

\begin{figure}[h]
 \centering
 \includegraphics[width=1.10\textwidth]{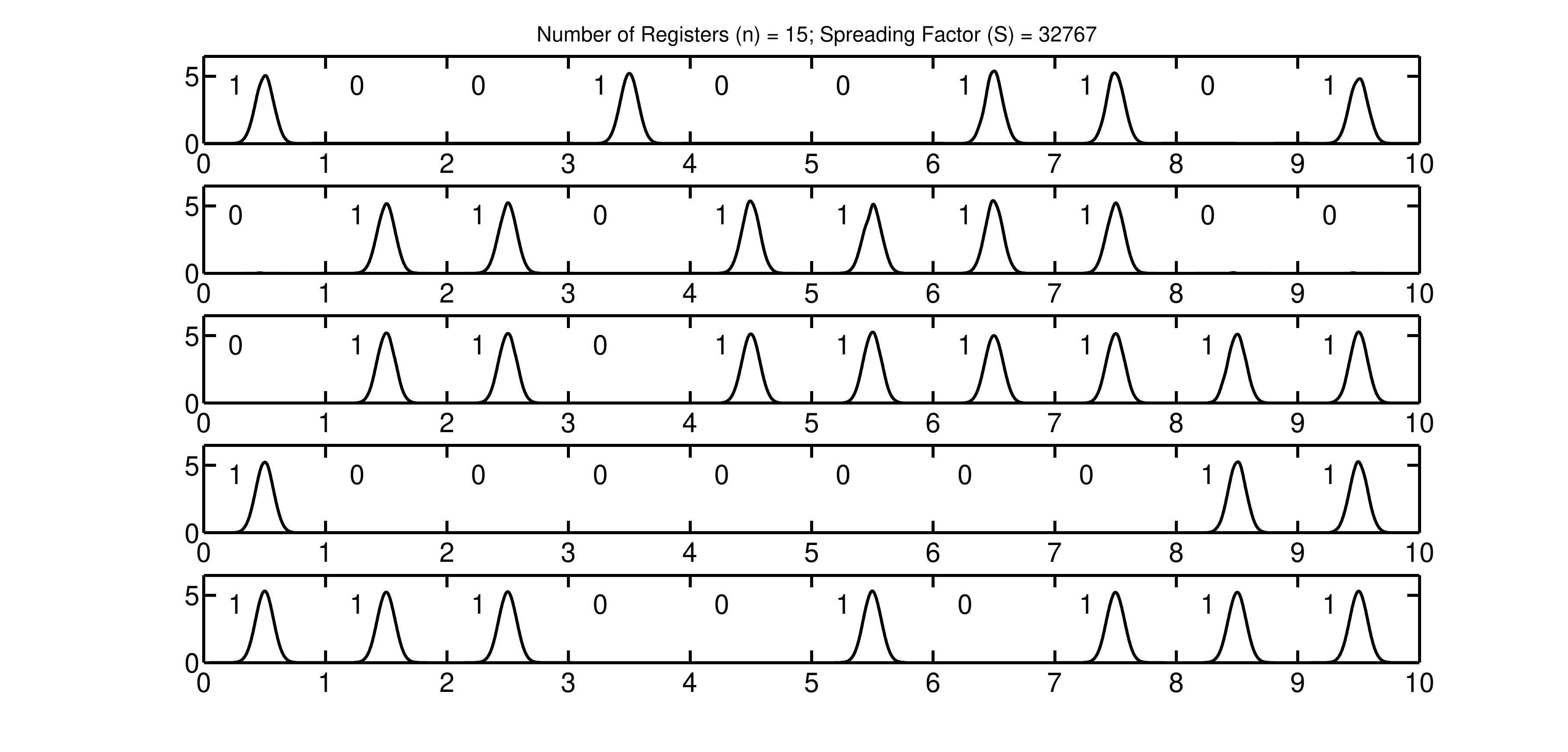}
\caption[1]{Spreading factor, $S$ = $2^{15}$ - 1 (with a code from an LFSR (Linear Feedback Shift Register) and n = 15 registers) for five users. The figure shows the output in terms of the density of the average photon number. Here, we obtain perfect output without any distortion and losses. In an ideal system, the represented output corresponds to the probability density squared $|\phi(t)|^2$.} \label{n81} 
 \end{figure}

In our analysis, we are considering five users who transmit eight bit random sequence, in which empty time bin is denoted by $0$ , and one photon Gaussian pulse is represented by  $1$. Here, we are considering two different values of the spreading factor $S = 2^{8} - 1$ and $S = 2^{15} - 1$, as shown in the Figs. \ref{n80}  and \ref{n81}, respectively. The result is shown in the form of density of the average photon number. Moreover, the output is in the form of probability density squared $|\phi(t)|^{2}$. These outputs indicate the losses, probability of photon loss, i.e. less than one information carrier (photon) in a time bin, and crosstalk, i.e. probability of finding a photon in an empty time bin, or more photons in a time bin, instead of one.\newline

These problems are addressed in Fig. \ref{n81}, where pulse distortion occurs during multiplexing operation. The pulses of different heights are indicated to represent the losses during transmission. Other than these, the residual pulses are indicated by 0, in which some portion of the information carrier photons in adjacent channels reach a user that must receive zero photons.\newline

\begin{figure}[h]
	  	       		    \centering
	  	       		     \includegraphics[width=1.10\textwidth]{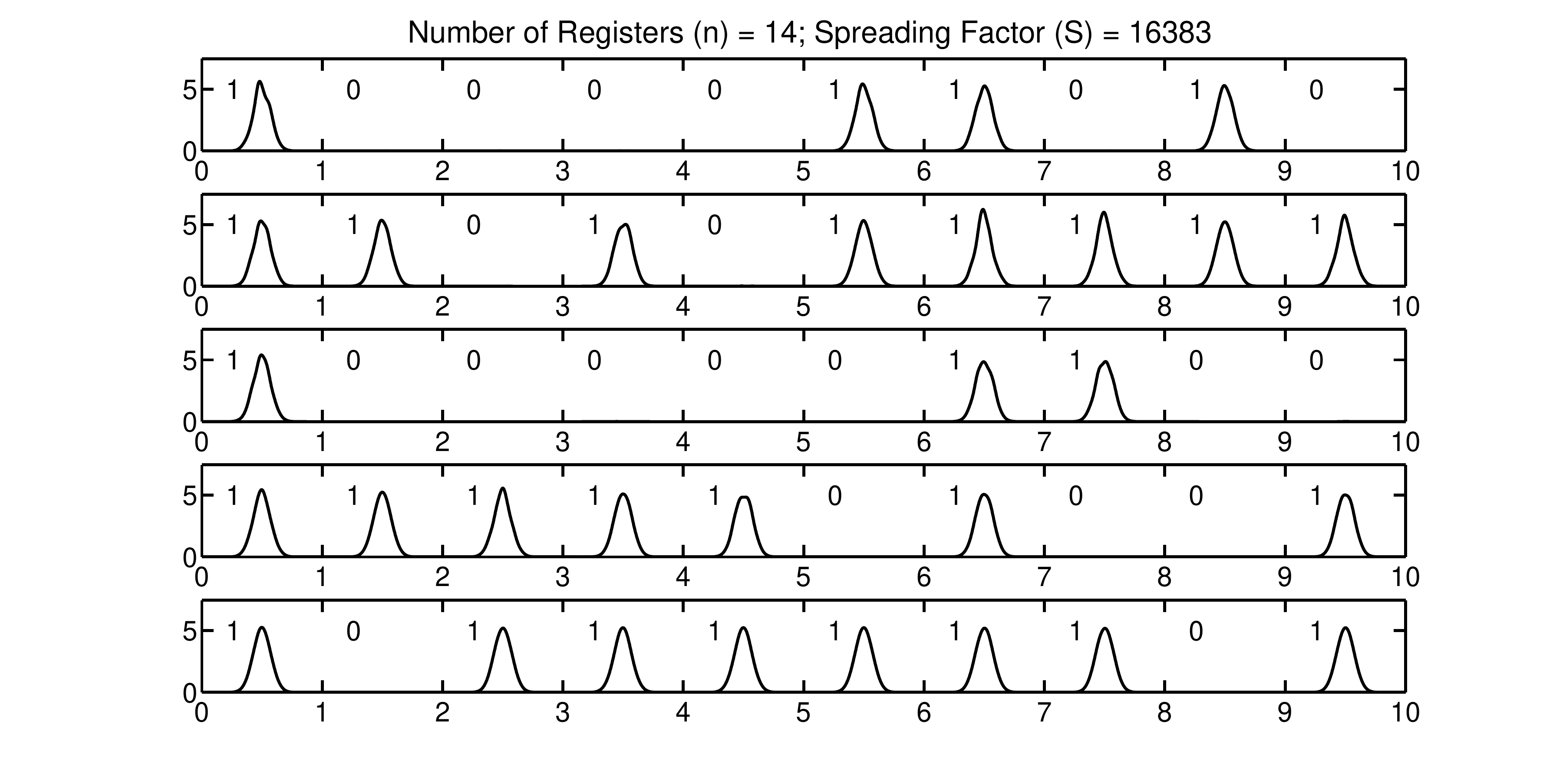}

	  	       				  	       		  	    	    	  	  	    \caption[1]{Example of transmission in a system with $S$ = $2^{14}$ - 1 (with a code
	  	       				  	       		  	    	    	  	  	    from an LFSR with n = 14 registers) and five users.} \label{n83} 
 		  	    	    	  	  	    	    	  	 	    	    	  	  	    	    	    \end{figure}

We have shown the improved result without any losses and crosstalk, as shown in   Figs. \ref{n81} and \ref{n83}. These results show that a perfect high value of spreading factor, $S$, eliminates the various losses described earlier, and correspondingly the density of the average photon number, particularly in this case,  is close to $|\phi(t)|^{2}$. In this case, the integration of the Gaussian pulse (area under the curve) is one.\newline

\begin{figure}[h]
	  	       		    \centering
	  	       		     \includegraphics[width=1.10\textwidth]{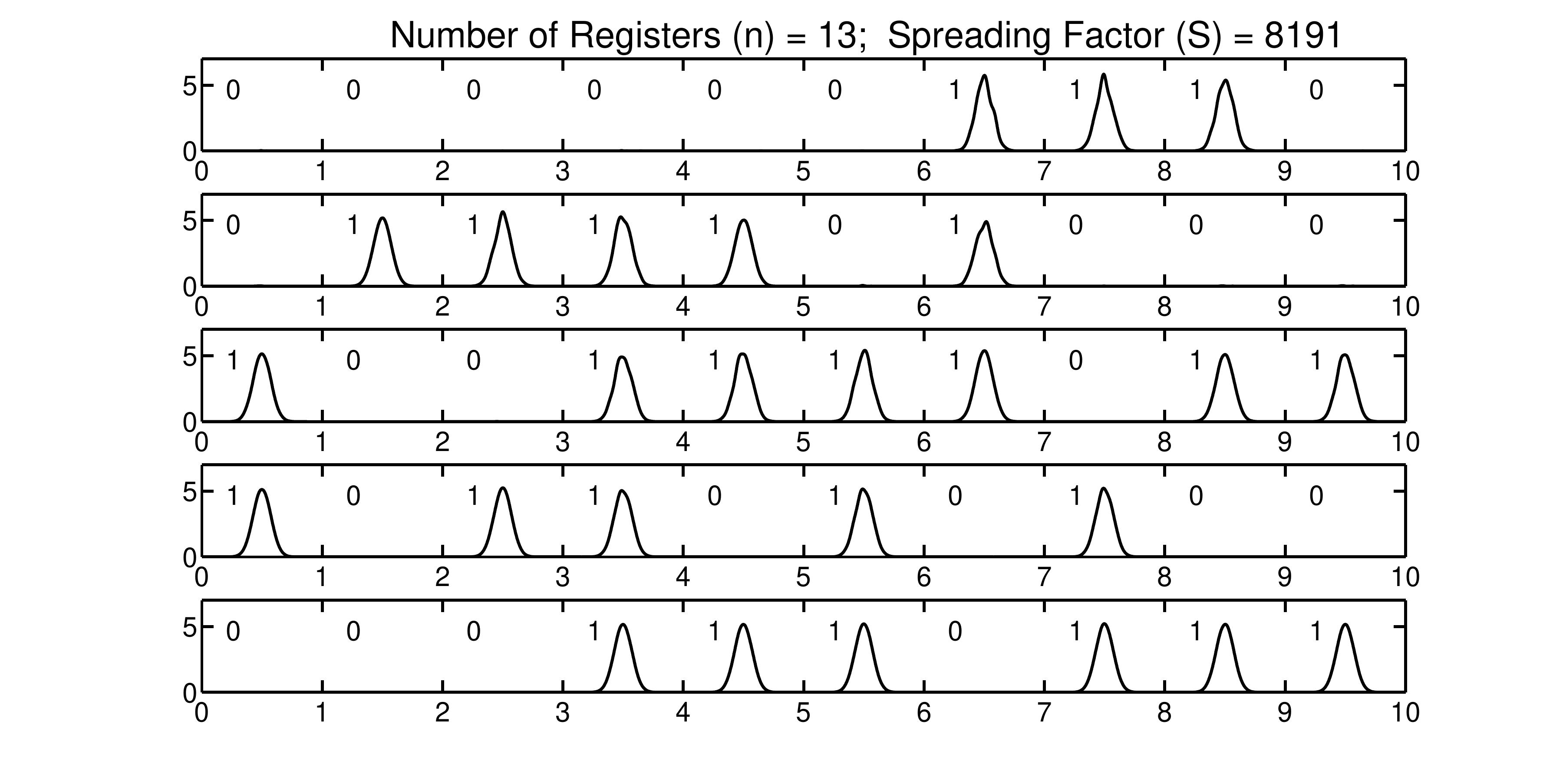}

	  	       				  	       		  	    	    	  	  	    \caption[1]{Example of transmission in a system with $S$ = $2^{13}$ - 1 (with a code
	  	       				  	       		  	    	    	  	  	    from an LFSR with n = 13 registers) and five users.} \label{n82} 
 		  	    	    	  	  	    	    	  	 	    	    	  	  	    	    	    \end{figure}

The additional results are shown to compare the effect of spreading factor, $S$ and number of users, as depicted in Fig. \ref{n82} . These parameters help to compare the various losses or the photon transmission in a wrong channel.\newline 
\begingroup
	  	    \setlength{\tabcolsep}{1pt} 
	  	    \renewcommand{\arraystretch}{2.50} 
	  	  \begin{table*}[ht]
	  	    \caption{ Photon Loss Probability}
	  	  	    \centering
	  	    	   
	  	    	    \begin{tabular}{|c| c| c| c| c}
	  	    	    \hline
	  	    	     Value of Spreading Factor & Five Users  & Twenty Users  & Fifty Users \\
	  	    	    	  	 \hline
	  	    	    S = $2^{8}$ - 1 & 0.3237 &0.8300  &0.9890  &  \\
	  	    	    \hline
	  	    	      S = $2^{10}$ - 1 & 0.1197 &0.3720  &0.6727  &  \\
	  	    	      	  	    	    \hline
	  	    	     S = $2^{12}$ - 1 & 0.0583 &0.1337  &0.2640  &  \\
	  	    	     	  	    	      	  	    	    \hline
	  	    	       S = $2^{14}$ - 1 & 0.0424 &0.0618  &0.0996  &  \\
	  	    	      	  	    	     	  	    	      	 [1ex]
	  	    	    \hline
	  	    	    \end{tabular}
	  	    	    \label{table:nonliin2}
	  	    	    \end{table*}
	   				\endgroup

\begingroup
	  	    \setlength{\tabcolsep}{1pt} 
	  	    \renewcommand{\arraystretch}{2.50} 
	  	  \begin{table*}[ht]
	  	    \caption{ Probability of Crosstalk}
	  	  	    \centering
	  	    	   
	  	    	    \begin{tabular}{|c| c| c| c| c}
	  	    	    \hline
	  	    	     Value of Spreading Factor & Five Users  & Twenty Users  & Fifty Users \\
	  	    	    	  	 \hline
	  	    	    S = $2^{8}$ - 1 & 0.0632 &0.2240  &0.3888  &  \\
	  	    	    \hline
	  	    	      S = $2^{10}$ - 1 & 0.0183 &0.0728  &0.1677  &  \\
	  	    	      	  	    	    \hline
	  	    	     S = $2^{12}$ - 1 & 0.0041 &0.0184  &0.0480  &  \\
	  	    	     	  	    	      	  	    	    \hline
	  	    	       S = $2^{14}$ - 1 & 0.0010 &0.0050  &0.0124  &  \\
	  	    	      	  	    	     	  	    	      	 [1ex]
	  	    	    \hline
	  	    	    \end{tabular}
	  	    	    \label{table:nonliin3}
	  	    	    \end{table*}
	   				\endgroup

\begingroup
	  	    \setlength{\tabcolsep}{1pt} 
	  	    \renewcommand{\arraystretch}{2.50} 
	  	  \begin{table*}[ht]
	  	    \caption{ Fidelity values for different users}
	  	  	    \centering
	  	    	   
	  	    	    \begin{tabular}{|c| c| c| c| c}
	  	    	    \hline
	  	    	     Single Photon as an information carrier & Five Users  & Twenty Users  & Fifty Users \\
	  	    	    	  	 \hline
	  	    	    $|0\rangle$ & 1.077.$10^{-3}$ & 2.330.$10^{-3}$  & 5.624.$10^{-3}$  &  \\
	  	    	    \hline
	  	    	      $|1\rangle$  & 1.077.$10^{-3}$ & 2.330.$10^{-3}$  & 5.620.$10^{-3}$   &  \\
	  	    	      	  	    	    \hline
	  	    	     $|+\rangle$ & 1.077.$10^{-3}$ & 2.340.$10^{-3}$   & 5.630.$10^{-3}$   &  \\
	  	    	     	  	    	      	  	    	    \hline
	  	    	       $|-\rangle$& 1.076.$10^{-3}$ & 2.332.$10^{-3}$  & 5.605.$10^{-3}$   &  \\
	  	    	      	  	    	     	  	    	      	 [1ex]
	  	    	    \hline
	  	    	    \end{tabular}
	  	    	    \label{table:nonliin4}
	  	    	    \end{table*}
	   				\endgroup  

Here we have computed probability of photon loss in the cases where only one channel transmits a Gaussian single photon pulse. This channel selection is randomly allotted. The results have been shown in Table \ref{table:nonliin2}, representing average photon loss probability for 200 different tests. The values depicted in the Table \ref{table:nonliin2}  are computed from the probability density, which denotes the availability of the photon in its original channel. \newline

It is observed that losses becomes effective in the communication system, as the number of users grow. Each multiplexer and demultiplexer stage introduces additional losses. Larger value of spreading factor, helps in filtering process, and filtering becomes more selective for higher values of $S$. Thus, more the spreading factor $S$, more the photon spread, hence larger bandwidth results in less losses and making it more acceptable for practical applications.\newline

Other source of error is crosstalk. Here, in such case, we have calculated the photon probability appearing in a channel which is empty. The transmitted single photon is random in phase. In these calculations, we randomly allocated empty channel and investigated the photon availability at the output. The area under the Gaussian photon pulse (integrating average photon number density in the considered channel), gives the value of the crosstalk for 128 runs where each user transmits 8 bits, as shown in Table \ref{table:nonliin3}. The considered channel should be empty in such cases, while calculating crosstalk.\newline

As the number of users increase, losses (crosstalk and probability of photon loss), also increase. On the other side, to avoid channel interference, we need to select longer codes. To make it practically acceptable and feasible in a realistic scenario, there must be fine tuning between the number of users and code selection.\newline

\begin{figure}[h]
	  	       		    \centering
	  	       		     \includegraphics[width=0.810\textwidth]{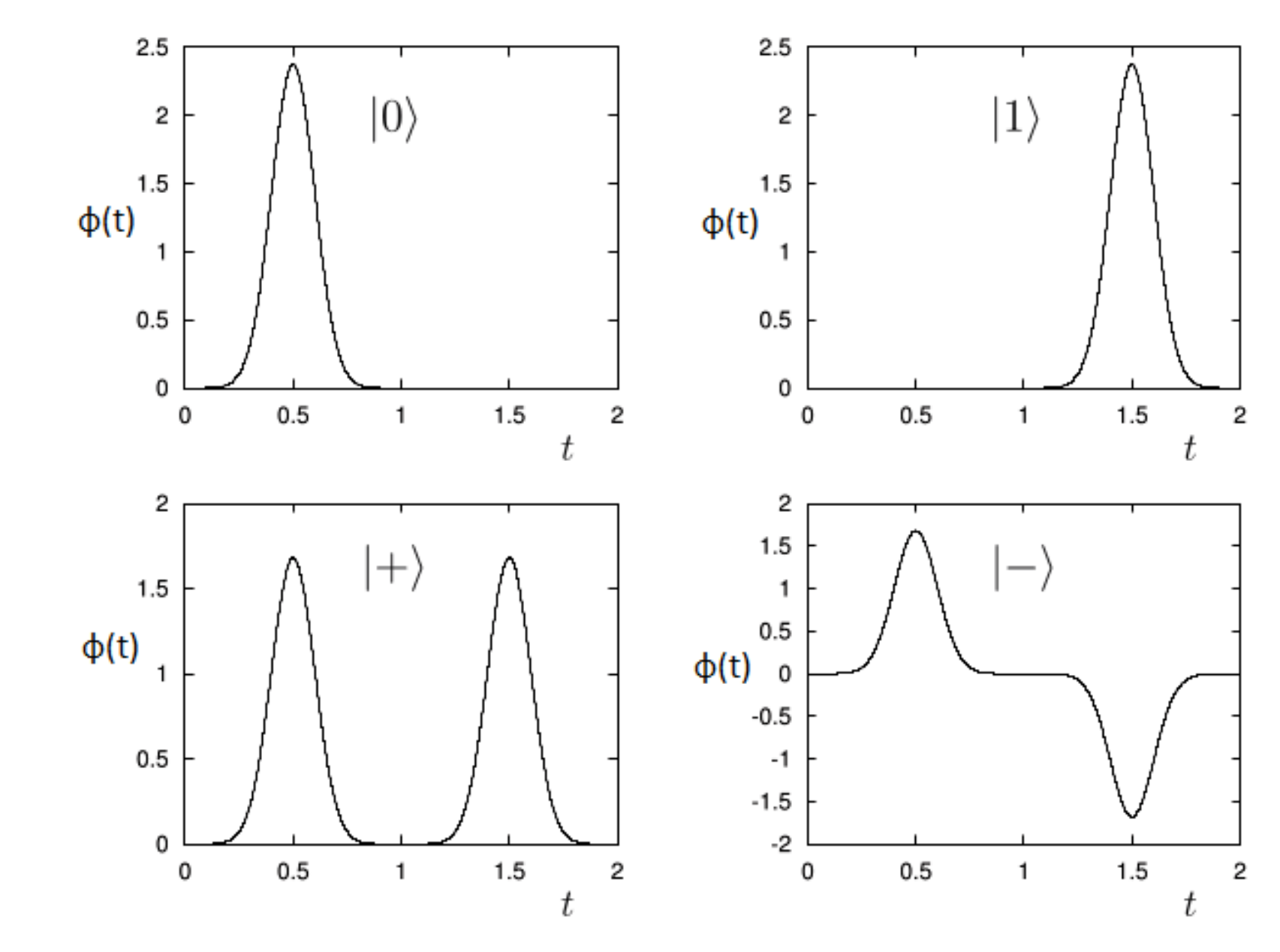}

	  	       				  	       		  	    	    	  	  	    \caption[1]{Waveform of coherent one-way Quantum Key Distribution (COW QKD).} \label{n88} 
 		  	    	    	  	  	    	    	  	 	    	    	  	  	    	    	    \end{figure}

It is necessary for a quantum system that it should follow principal of quantum superposition. Any quantum system used in quantum communication applications must obey this property of quantum superposition, as required in coherent one-way (COW) QKD protocol \cite{walenta2014fast, stucki2005fast, sibson2017chip}. In the present study, we are considering four different quantum states $|0\rangle$, $|1\rangle$, $|+\rangle$, and $|-\rangle$ which serve as single photon information carrier. We have defined these quantum states, and  shown in Fig. \ref{n88}. A state with photon present in the first time bin, but empty in the second time bin, is denoted by $|0\rangle$. While if the photon is present in the second time bin, but empty in the first time bin, is represented by the state $|1\rangle$. Here each of the time bins is of duration $T$. Following the same pattern, the superposition of the quantum states, $|0\rangle$ and $|1\rangle$ are written as $|+\rangle$ ( $|+\rangle = \frac{|0\rangle + |1\rangle}{\sqrt 2})$ and ( $|-\rangle = \frac{|0\rangle - |1\rangle}{\sqrt 2})$. These superposition states span the time duration from $0$ to $2T$, as shown in Fig. \ref{n88}.\newline

Let $\phi(t)$ is the initial wavefunction, which travels through the optical channel. At the receiver end, we get a distorted wavefunction $\widetilde{\phi}(t)$. Fidelity,  $F$,  is used to compare between the input and output states. It can be defined from overlap integral

\begin{eqnarray}
F = |\int\widetilde{\phi^{*}}(t)\phi(t)dt|^{2}
\end{eqnarray}

 where $\phi(t)$ denotes the original photon's wavefunction at the input side, and $\widetilde{\phi^{*}}(t)$ is the complex conjugate of the distorted wavefunction at the receiver end. The asterik ($*$) indicates the complex conjugate. Here we have not taken the effect of the time of flight, which is the photons travel time taken through the optical network. \newline

The average fidelity is computed between the input and the normalized output state. The four quantum states with different number of users and their corresponding fidelity values are calculated and shown in Table \ref{table:nonliin4}. These values are written for spreading factor, $S = 2^{10} - 1$. All fidelity values are close to 1.\newline

\section{DISCUSSION AND CONCLUSION}
Considering a single optical fiber as a channel, we studied spread spectrum technique, in which multiple users transmit their qubits. A unique code is deployed for each user to spread the original message signal. The original message signal is obtained by considering the wavefunction of each photon. The unique advantage of the spread spectrum approach is that, this technology offers extended bandwidth. For example, CDMA (code division multiple access) is a type of spread-spectrum technique, where $N$ different qubits are encoded in $N$ different photons and only a single optical fiber is used as a transmission channel.\newline

During successive multiplexing stages, a new qubit is added in the channel corresponds to each new user. The photon loss probability of each ``old" photon is $\frac{1}{S}$. This is similar to that of demultiplexing operation. There are 
$2N - 2$  lossy stages for a photon while passing from various multiplexing-demultiplexing stages. Here $N$ represents the number of users sharing a particular channel. The maximum probability of photon loss is $\frac{2N-2}{S}$, which can be reduced further by the proper selection of addition and extraction of qubits of each user. The Figs. \ref{mux} and \ref{demux} will add coupling losses that should in principle be taken into account.\newline

Continuous-variable (CV) quantum key distribution is different from the standard quantum key distribution method  for detecting weak optical signals \cite{hirano2003quantum, grosshans2003quantum, jouguet2013experimental, wang201525, huang2016long, imre2012advanced, hanzo2012wireless}. Here, optical fiber channel is used as a transmission medium, in which codes are deployed from code division multiple access. One very significant parameter is spreading factor $S$, which helps in reducing noise effects on system performance, and at the same time allow more users to share the channel. For adequate separation,  $N\leq S$, is the condition, which gives appropriate number of orthogonal codes. The acceptable experimental value of $S$ is $2^{15}-1$, which is practically possible without further overlap \cite{belthangady2010hiding}. There are practical limits on modulators. Modulation rate  of 10--100 Gbps is achieved. Exceeding the modulation rate, enhanced smooth transition occur in the codes. These smooth transitions are undesirable and degrade the overall performance of the multiplexing stages. These effects lead to the limits on the modulation rates and code length. Photons with spreading factor values; $S \equiv 2^{13} - 1$ or $2^{14}-1$, and with wavefunctions in microsecond length (photons in MHz range) is practically possible.\newline

Coupling losses at the optical elements are also likely to be a major limitation. In many QKD networks, losses limit the maximum communication distance, but spreading provides certain protection against noise that reduce the strength of losses on the signal-to-noise ratio of the data link. The noise that has been picked up in the channel \cite{sharma2018decoherence} is spread at the receiver and the filter that rejects adjacent channels also stops a greater proportion of the energy of the noise. This is an independent effect of spreading and can be used to extend the reach of QKD links with a single photon. In that case, the photon needs not to be spread with a modulator. An interesting alternative is using spread spectral teleportation, a teleportation protocol that can stretch or shrink the wavefunction in frequency \cite{humble2010spectral}. This kind of teleportation could extend the applicability of spread spectrum methods to quantum repeater networks \cite{briegel1998quantum}.\newline

We have discussed that there is a trade off between spreading factor $S$, and the number of users, $N$. The only solution of this problem is that, to avoid the losses, we can restrict the number of users to access the channel during communication, which results in lower losses ($\frac{1}{S}$ ).\newline

The deployment of integrated optical elements, such as integrated microring structures with discrete elements are a substitute  of  fiber Bragg-grating and the circulators. These microring structures  perform as frequency selective filters and routing devices, thereby eliminating the requirement of amplifiers and giving birth of energy-efficient quantum optical code division multiple access \cite{little1998ultra, xiao2008silicon}. \newline

The present study of the multiplexing method can also be used to combine classical and quantum data. Most QKD networks send photons through what are called dark fibers, which are reserved for quantum use and carry no classical data. Classical and quantum information channels can share the same fiber if they are assigned different frequency bands, but Raman scattering and other processes triggered by the classical optical signal introduce noise into the photon channel. The present study of add-drop architecture offers a new way to introduce a single photon into an optical fiber that carries classical signals. The method allows insertion in already deployed optical networks and spreading helps to fight the noise in the communication channel.\newline

 \section*{Acknowledgements}

\end{document}